\documentclass[aps,showpacs,preprintnumbers,amsmath, amssymb]{revtex4}

\usepackage{float}
\usepackage{graphics,epsfig}
\usepackage{graphicx}
\usepackage{dcolumn}
\usepackage{bm}

\begin{document}
\baselineskip=0.8 cm

\title{{\bf No hair theorem for massless scalar fields outside asymptotically flat horizonless reflecting compact stars}}
\author{Yan Peng$^{1}$\footnote{yanpengphy@163.com}}
\affiliation{\\$^{1}$ School of Mathematical Sciences, Qufu Normal University, Qufu, Shandong 273165, China}

\vspace*{0.2cm}
\begin{abstract}
\baselineskip=0.6 cm
\begin{center}
{\bf Abstract}
\end{center}

In a recent paper, Hod started a study on no scalar hair theorem
for asymptotically flat spherically symmetric neutral horizonless reflecting compact stars.
In fact, Hod's approach only rules out massive scalar fields.
In the present paper, for massless scalar fields outside neutral horizonless reflecting compact stars,
we provide a rigorous mathematical proof on no hair theorem.
We show that asymptotically flat spherically symmetric neutral horizonless reflecting compact stars cannot
support exterior massless scalar field hairs.

\end{abstract}

\maketitle
\newpage
\vspace*{0.2cm}

\section{Introduction}

Recently, the first ever image of a black hole has been captured by a network
of eight radio telescopes around the world \cite{first}. These discoveries open up hope
to directly test various black hole theories from astronomical aspects.
One remarkable property of classical black holes is the
famous no hair theorem \cite{Bekenstein}-\cite{JBN}.
If generically true, such no hair theorem would signify
that asymptotically flat black holes cannot support scalars, massive
vectors and Abelian Higgs hairs in exterior regions,
for recent references see \cite{mr1}-\cite{CW} and reviews see \cite{Bekenstein-1,CAR}.
It was believed that no hair behaviors are due to the
existence of one-way absorbing horizons.

However, it was recently found that no hair behavior also appears in the
background of horizonless reflecting compact stars.
In the asymptotically flat gravity, Hod firstly proved no static scalar hair theorem
for neutral horizonless reflecting compact stars \cite{Hod-6}.
In the asymptotically dS gravity, it was found that
neutral horizonless reflecting compact stars
cannot support the existence of massive scalar, vector and tensor field
hairs \cite{Bhattacharjee}.
When considering a charged background,
large reflecting shells can exclude
static scalar field hairs \cite{Hod-8,Hod-9,Yan Peng-1}.
Similarly, static scalar field hairs cannot
exist outside charged reflecting compact stars
of large size \cite{Hod-10,Yan Peng-2,Yan Peng-3,Yan Peng-4,Yan Peng-5}.
With field-curvature couplings, such no hair theorem could
also hold in the horizonless gravity \cite{nonm1,nonm2,nonm3}.
Moreover, we proved no hair theorem for horizonless compact stars
with other boundary conditions \cite{Yan Peng-6,Yan Peng-7,Yan Peng-8}.

As is well known, scalar field mass usually plays an important role
in the scalar hair formation.
For massless scalar fields $\psi(r)$,
no scalar hair theorem was investigated
in the background of horizonless compact stars \cite{Hod-6},
where the relation $\psi(r_{peak})\psi''(r_{peak})<0$
at the extremum point $r=r_{peak}$ is essential in Hod's present proof.
However, the general characteristic relation
at the extremum point should be $\psi(r_{peak})\psi''(r_{peak})\leqslant 0$
and in fact, $\psi''(r_{peak})=0$ holds for some solutions.
So Hod's approach only ruled out massless scalar fields with
$\psi''(r_{peak})\neq 0$ and nontrivial solutions with $\psi''(r_{peak})=0$
cannot be excluded. Then it is of some importance to search for a
mathematical proof on no hair theorem
for massless scalar field hairs.

In the following, we consider static massless scalar fields in the
background of asymptotically flat spherical neutral horizonless reflecting compact stars.
We provide a rigorous mathematical proof on no hair theorem
for massless scalar fields.
We summarize main results in the last section.

\section{No massless scalar hair for horizonless reflecting compact stars}

We consider massless scalar fields outside
asymptotically flat horizonless compact stars.
We define the radial coordinate $r=r_{s}$ as the star radius.
And the curved spherically symmetric spacetime is described by line element \cite{nonm1,nonm2}
\begin{eqnarray}\label{AdSBH}
ds^{2}&=&-e^{\nu}dt^{2}+e^{\lambda}dr^{2}+r^{2}(d\theta^2+sin^{2}\theta d\phi^{2}),
\end{eqnarray}
where $\nu=\nu(r)$ and $\lambda=\lambda(r)$.
Asymptotic flatness of the spacetime requires infinity behaviors
$\nu(r\rightarrow \infty)\backsim O(r^{-1})$
and $\lambda(r\rightarrow \infty)\backsim O(r^{-1})$.

The Lagrange density with massless scalar fields
minimally coupled to gravity is given by
\begin{eqnarray}\label{lagrange-1}
\mathcal{L}=R-(\partial_{\alpha} \psi)^{2}.
\end{eqnarray}
Here $\psi(r)$ is the scalar field
and R is the scalar Ricci curvature of the spacetime.

From the metric (1) and the action (2), we obtain the
scalar field equation
\begin{eqnarray}\label{BHg}
\psi''+(\frac{2}{r}+\frac{\nu'}{2}-\frac{\lambda'}{2})\psi'=0.
\end{eqnarray}
Around the infinity, the scalar field equation
can be approximated by
\begin{eqnarray}\label{BHg}
\psi''+\frac{2}{r}\psi'=0.
\end{eqnarray}
It yields the infinity boundary behavior
\begin{eqnarray}\label{AdSBH}
\psi\sim A+\frac{B}{r},
\end{eqnarray}
where A and B are integral constants.
The finite ADM mass condition yields that
the static scalar field must asymptotically approaches
zero at the infinity. So we fix $A=0$. Then
the scalar field satisfies the infinity boundary condition
\begin{eqnarray}\label{AdSBH}
\psi(r\rightarrow\infty)=0.
\end{eqnarray}

At the star surface, we impose the reflecting boundary condition
\begin{eqnarray}\label{AdSBH}
\psi(r_{s})=0.
\end{eqnarray}

According to relations $\psi(r_{s})=0$
and $\psi(r\rightarrow\infty)=0$,
$\psi(r)$ must possesses one extremum point $r_{peak}$,
which is between the star surface $r_{s}$ and the infinity.
With the symmetry $\psi\rightarrow -\psi$ of equation (3),
it is enough for us to only consider the case of positive local maximum value in the following.

Metric solutions were assumed to be real analytic
in the proof of the uniqueness of the Kerr solution \cite{BCU,DRU,Hawking}.
A real function is analytic if it can be locally expressed with Taylor series.
In this work, we also assume that the metric solutions are real analytic.
Then a positive constant $\delta$ exists and
in the range $(r_{peak}-\delta,r_{peak}+\delta)$,
metric solutions can be expanded as
\begin{eqnarray}\label{BHg}
\nu(r)=\sum_{n=0}^{\infty}a_{n}(r-r_{peak})^{n},
\end{eqnarray}
\begin{eqnarray}\label{BHg}
\lambda(r)=\sum_{n=0}^{\infty}b_{n}(r-r_{peak})^{n},
\end{eqnarray}
where $a_{n}=\frac{\nu^{(n)}(r_{peak})}{n!}$ and $b_{n}=\frac{\lambda^{(n)}(r_{peak})}{n!}$.
Solutions $\psi(r)$ of equation (3) are real analytic
according to Cauchy-Kowalevski theorem, which states that
solutions of differential equations
are real analytic when coefficients are real analytic \cite{BCU,DRU,Hawking,NAM}.
In the same range $(r_{peak}-\delta,r_{peak}+\delta)$,
the scalar field can be expressed as
\begin{eqnarray}\label{BHg}
\psi(r)=\sum_{n=0}^{\infty}c_{n}(r-r_{peak})^{n}
\end{eqnarray}
with $c_{n}=\frac{\psi^{(n)}(r_{peak})}{n!}$,
which can be obtained by putting (8) and (9) into (3) and considering
terms order by order.

Now we show that a nonzero $c_{n}$ ($n\geqslant 1$) should exist for nontrivial solution $\psi(r)$.
Otherwise, $\psi(r)$ is a constant in the range $(r_{peak}-\delta,r_{peak}+\delta)$.
Then we can search for the largest R where $\psi(r)$ is a constant in the
range $(r_{peak}-\delta,R]$. Since $\psi(r)$ is real analytic, there
is a constant $\tilde{\delta}>0$ and in the range $(R-\tilde{\delta},R+\tilde{\delta})$,
the scalar field can be expressed as
\begin{eqnarray}\label{BHg}
\psi(r)=\sum_{n=0}^{\infty}d_{n}(r-R)^{n}.
\end{eqnarray}
If we approach R from the left side, we find all coefficients $d_{n}=0$ for $n\geqslant 1$
since $\psi(r)$ is a constant in the range $(R-\tilde{\delta},R]$.
However, if we approach R from the right side, a nonzero
$d_{n}\neq 0$ with $n\geqslant 1$ should exist since
$\psi(r)$ is not a constant in the range $[R,R+\tilde{\delta})$.
This contradiction leads to the conclusion that
a nonzero $c_{n}=\frac{\psi^{(n)}(r_{peak})}{n!}$ ($n\geqslant 1$) exists.

By considering leading terms of (10), we obtain following conclusions.

(I)Firstly, there is $\psi'(r_{peak})=0$. Otherwise,
$\psi(r)=\psi(r_{peak})+\psi'(r_{peak})(r-r_{peak})+\ldots$ and $\psi(r)$ cannot
has extremum value at the point $r_{peak}$.

(II)In the case of $\psi''(r_{peak})\neq 0$, we will have $\psi''(r_{peak})< 0$. Otherwise,
$\psi$ cannot has local maximum extremum value at the point $r_{peak}$
since $\psi(r)=\psi(r_{peak})+\frac{\psi''(r_{peak})}{2}(r-r_{peak})^2+\ldots$.
In this work, we only consider the case of positive local maximum value according
to the symmetry $\psi\rightarrow -\psi$ of equation (3).

(III)In the case of $\psi''(r_{peak})= 0$, we will have $\psi^{(3)}(r_{peak})= 0$. Otherwise,
$\psi$ cannot have local maximum extremum value at the point $r_{peak}$ since
$\psi(r)=\psi(r_{peak})+\frac{\psi'''(r_{peak})}{3}(r-r_{peak})^3+\ldots$.

(IV)In the case of $\psi''(r_{peak})= 0$ and $\psi^{(4)}(r_{peak})\neq0$,
we will have to impose $\psi^{(4)}(r_{peak})< 0$ to obtain
a local maximum extremum value for $\psi$ at the point $r_{peak}$.
In this case, there is the relation
$\psi(r)=\psi(r_{peak})+\frac{\psi^{(4)}(r_{peak})}{24}(r-r_{peak})^4+\ldots$.

Following this analysis, we can obtain an even number
N, which satisfies
\begin{eqnarray}\label{BHg}
\psi'=0,~\psi''=0,~\psi^{(3)}=0,~\psi^{(4)}=0,\ldots,\psi^{(N-1)}=0,~\psi^{(N)}<0~~~for~~~r=r_{peak}.
\end{eqnarray}
At the extremum point $r=r_{peak}$, equation (3) yields relations

We define a new function $f(r)=\frac{2}{r}+\frac{\nu'}{2}-\frac{\lambda'}{2}$.
Then the equation (3) can be expressed as
\begin{gather}
\psi''+f\psi'=0.\tag{13}
\end{gather}
As we stated, metric solutions are assumed to be real analytic,
which is the same as cases in the proof of the uniqueness of the Kerr solution \cite{BCU,DRU,Hawking}.
According to Cauchy-Kowalevski theorem, as we have shown, equation (3) possesses a locally real analytic
solution $\psi(r)$ around $r_{peak}$. Since metric functions and $\psi(r)$ are real analytic
around $r_{peak}$, both $f'(r_{peak}),f''(r_{peak}),f'''(r_{peak}),f^{(4)}(r_{peak}),\ldots$
and $\psi'(r_{peak}),\psi''(r_{peak}), \psi'''(r_{peak}),\psi^{(4)}(r_{peak}),\ldots$ exist.
Taking the derivative of both sides of the equation (13),
we get the equation
\begin{gather}
(\psi''+f\psi')'=0,\tag{14}
\end{gather}
which holds round $r_{peak}$.
The relation (14) is equal to
\begin{gather}
\psi'''+f\psi''+f'\psi'=0.\tag{15}
\end{gather}
From (15), we obtain the equation
\begin{gather}
(\psi'''+f\psi''+f'\psi')'=0.\tag{16}
\end{gather}
The relation (16) can be transformed into
\begin{gather}
\psi^{(4)}+f\psi'''+2f'\psi''+f''\psi'=0.\tag{17}
\end{gather}
Along this line, we can obtain the following relation
\begin{gather}
\psi^{(N)}+f\psi^{(N-1)}+(N-2)f'\psi^{(N-2)}+\ldots+(N-2)f^{(N-3)}\psi''+f^{(N-2)}\psi'=0.\tag{18}
\end{gather}

At the extremum point, relations (12) are in contradiction with the relation (18).
Due to this contradiction, there is no nontrivial scalar field solution of equation (3).
We conclude that asymptotically flat spherical neutral horizonless reflecting compact stars cannot support
exterior massless scalar field hairs.

\section{Conclusions}

We studied no hair theorem for static massless scalar fields
outside the asymptotically flat spherically symmetric horizonless reflecting compact stars.
We obtained the characteristic relations (12) at extremum points,
which are in contradiction with the equation (18).
That is to say there is no nontrivial scalar field solution of equation (3).
We concluded that asymptotically flat spherically symmetric horizonless reflecting
compact stars cannot support the existence of exterior massless scalar fields.
In this work, we provided a rigorous mathematical proof on no hair theorem for massless scalar fields.

\begin{acknowledgments}

This work was supported by the Shandong Provincial Natural Science Foundation of China under Grant
No. ZR2018QA008. This work was also supported by a grant from Qufu Normal University
of China under Grant No. xkjjc201906.

\end{acknowledgments}

\end{document}